\documentclass[
    aip,
    reprint,
    superscriptaddress,
    amsmath,
    amssymb,
    amssymb,
    amsfonts
]{revtex4-2}

\usepackage{dcolumn}
\usepackage{bm}
\usepackage{hyperref}
\usepackage{comment}
\usepackage{xcolor}
\usepackage{graphicx}
\usepackage{layouts} 
\usepackage{tikz}
\usepackage{pgfplots}
\usepackage{ifthen}
\usepackage{babel} 
\usetikzlibrary{positioning, calc}
\usetikzlibrary{decorations.pathmorphing} 
\usetikzlibrary{arrows.meta} 
\usepackage[export]{adjustbox}
\usepackage{xspace} 
\usepackage{pgf}
\usepackage{bbold}
\usepackage{soul}
\usepackage{lmodern}
\usepackage{placeins}
\usepackage{mathrsfs}
\usepackage[inline]{enumitem}
\usepackage{booktabs}
\usepackage{multirow}
\usepackage[normalem]{ulem}

\newcommand{\br}[1]{\left( #1 \right)}
\newcommand{\brr}[1]{\left[ #1 \right]}

\newcommand{\tableheadline}[1]{\multicolumn{1}{c}{\bfseries #1}}
\newcommand{\mathdefault}[1][]{} 
\newcommand{\avg}[1]{\langle {#1} \rangle}
\newcommand{\abs}[1]{\vert {#1} \vert}

\newcommand{\ket}[1]{\vert {#1}\rangle}

\newcommand{\ketbra}[2]{\vert {#1} \rangle \langle{#2}\vert}
\newcommand{\matrixel}[3]{\langle #1 \vphantom{#3} | #2 | #3 \vphantom{#1} \rangle}

\DeclareMathOperator*{\argmax}{arg\,max}
\newcommand*{\ie}{i.e.\@\xspace}
\newcommand{\vect}[1]{\boldsymbol{#1}} 

\newcommand{\red}[1]{{\color[rgb]{0.7,0,0} #1}}

\makeatletter
\newcommand{\hrules}{%
\@for\i:=1,2,3,4,5\do{%
\noindent\red{\rule{\linewidth}{1pt}}\par\vspace{-1ex}%
}%
\vspace{1ex}
}
\makeatother

\newcommand{\Op}{\Omega_\mathrm{p}}
\newcommand{\Opmax}{\Omega_\mathrm{p}^\mathrm{max}}
\newcommand{\Os}{\Omega_\mathrm{s}}
\newcommand{\Osmax}{\Omega_\mathrm{s}^\mathrm{max}}

\newcommand{\Opsmax}{\Omega_\mathrm{p/s}^\mathrm{max}}
\newcommand{\ti}{t_\mathrm{i}}
\newcommand{\tf}{t_\mathrm{f}}

\newcommand{\de}{\mathrm{d}}
\newcommand{\xii}{\xi^{(r)}}

\newcommand{\sz}[1][]{\ifthenelse{\equal{#1}{}}{\sigma_z}{\sigma_z^{(#1)}}}
\newcommand{\sx}[1][]{\ifthenelse{\equal{#1}{}}{\sigma_z}{\sigma_x^{(#1)}}}
\newcommand{\sm}[1][]{\ifthenelse{\equal{#1}{}}{\sigma_-}{\sigma_-^{(#1)}}}
\renewcommand{\sp}[1][]{\ifthenelse{\equal{#1}{}}{\sigma_+}{\sigma_+^{(#1)}}}

\renewcommand{\sz}[1][]{\ifthenelse{\equal{#1}{}}{\sigma^z}{\sigma^z_{#1}}}
\renewcommand{\sx}[1][]{\ifthenelse{\equal{#1}{}}{\sigma^x}{\sigma^x_{#1}}}
\renewcommand{\sm}[1][]{\ifthenelse{\equal{#1}{}}{\sigma^-}{\sigma^-_{#1}}}
\renewcommand{\sp}[1][]{\ifthenelse{\equal{#1}{}}{\sigma^+}{\sigma^+_{#1}}}

\newcommand{\Hc}{H_\mathrm{c}}

\newcommand{\Hnoise}{H_\mathrm{n}}
\newcommand{\rhof}{\rho_\mathrm{f}}

\begin{document}

\title{Detection of noise correlations in two qubit systems by Machine Learning}

\author{Dario Fasone}
\thanks{These authors contributed equally to this work.}
\affiliation{Dottorato di Ricerca in ``Quantum Technologies'', Universit\`a di Napoli Federico II, Napoli, Italy}
\affiliation{Dipartimento di Fisica e Astronomia ``Ettore Majorana'', Università di Catania, Via S. Sofia 64, 95123 Catania, Italy}
\author{Shreyasi Mukherjee}
\thanks{These authors contributed equally to this work.}
\affiliation{Dipartimento di Fisica e Astronomia ``Ettore Majorana'', Università di Catania, Via S. Sofia 64, 95123 Catania, Italy}
\author{Dario Penna}
\affiliation{Leonardo S.p.A., Cyber \& Security Solutions, 95121, Catania, Italy}
\author{Fabio Cirinn\`a}
\affiliation{Leonardo S.p.A., Cyber \& Security Solutions, 95121, Catania, Italy}
\author{Mauro Paternostro}
\affiliation{Universit\`a degli Studi di Palermo, Dipartimento di Fisica e Chimica ``Emilio Segré'', via Archirafi 36, I-90123 Palermo, Italy}
\author{Elisabetta Paladino}
\affiliation{Dipartimento di Fisica e Astronomia ``Ettore Majorana'', Università di Catania, Via S. Sofia 64, 95123 Catania, Italy}
\affiliation{Istituto Nazionale di Fisica Nucleare, Sezione di Catania, 95123, Catania, Italy}
\affiliation{CNR-IMM, UoS Università, 95123, Catania, Italy}
\author{Luigi Giannelli}
\email[Corresponding author: ]{luigi.giannelli@dfa.unict.it}
\affiliation{Dipartimento di Fisica e Astronomia ``Ettore Majorana'', Università di Catania, Via S. Sofia 64, 95123 Catania, Italy}
\affiliation{Istituto Nazionale di Fisica Nucleare, Sezione di Catania, 95123, Catania, Italy}
\author{Giuseppe A. Falci}
\affiliation{Dipartimento di Fisica e Astronomia ``Ettore Majorana'', Università di Catania, Via S. Sofia 64, 95123 Catania, Italy}
\affiliation{Istituto Nazionale di Fisica Nucleare, Sezione di Catania, 95123, Catania, Italy}

\date{\today}

\begin{abstract}
    We introduce and validate a machine-learning assisted quantum sensing protocol
    to classify spatial and temporal correlations of classical noise affecting two
    ultrastrongly coupled qubits. We consider six distinct classes of Markovian
    and non-Markovian noise. Leveraging the sensitivity of a coherent population
    transfer protocol under three distinct driving conditions, the various forms
    of noise are discriminated by only measuring the final transfer efficiencies.
    Our approach achieves $\gtrsim 94\%$ accuracy in classification providing a near-perfect
    discrimination between Markovian and non-Markovian noise. The method
    requires minimal experimental resources, relying on a simple driving scheme providing
    three inputs to a shallow neural network with no need of measuring time-series
    data or real-time monitoring. The machine-learning data analysis acquires
    information from non-idealities of the coherent protocol highlighting how
    combining these techniques may significantly improve the characterization of
    quantum-hardware.
\end{abstract}

\maketitle

\section{\label{sec:Intro}Introduction}
Recently, accurate control of quantum systems has been achieved harnessing their
inherent coherence to perform tasks in quantum communication and computation~\cite{KochEQT2022quantum,AcinNJP2018quantum}.
However, the loss of fidelity of quantum operations is still a problem mainly because
of the interaction of the system with environmental degrees of freedom leading to
decoherence~\cite{ZurekRMP2003decoherence}. Therefore, methods for noise diagnostics
and strategies to mitigate its effects~\cite{LimPRAppl2025} are paramount for the advancement of quantum
technologies.

While for \emph{single} qubits, optimized design has allowed to achieve protection
from decoherence beyond the error-correction threshold~\cite{kjaergaard_superconducting_2020,postler_demonstration_2024},
substantial work has to be done in upscaled quantum architectures. In such contexts,
coherence times are shorter and detrimental effects of time- and space-correlated
noise may be significant. For example, in solid-state devices, impurities in the
substrates are sources of non-Markovian~\cite{MailePRB2024} and space-correlated noise~\cite{FalciPS2012effects}
that may also originate from shared control lines or collective electromagnetic modes.
In recent experiments, space-correlated 1/f noise in superconducting and semiconducting
spin qubits has been characterized \cite{von_lupke_two-qubit_2020,boter_spatial_2020,zou_spatially_2023,rojas-arias_spatial_2023,YonedaNP2023noisecorrelation}.
Similar correlations have been measured, long ago, in systems of small metallic
tunnel junctions \cite{ZorinPRB1996background}. 
The characterization of noise correlations is thus a critical requirement for the advancement of large-scale quantum computing, as standard protocols and quantum error correction (QEC) thresholds often assume negligible correlations between qubit errors. This assumption may not hold in upscaled quantum architectures~\cite{vepsalainen_impact_2020} and can require the development of new noise-mitigation strategies.
A variety of spectroscopic and computational methods have been developed to detect and classify these environmental influences. Experimental efforts have largely focused on spectroscopic protocols to quantify spatial dependencies. In Ref.~\cite{boter_spatial_2020} the authors utilized the coherence times of specific parallel and anti-parallel Bell states to distinguish between correlated and anti-correlated noise in Si/SiGe qubits. Building on this, interleaved Ramsey sequences and Bayesian estimation were implemented to reconstruct cross-power spectral densities~\cite{YonedaNP2023noisecorrelation,rojas-arias_spatial_2023}. To avoid the need for entangled-state preparation, Ref.~\cite{von_lupke_two-qubit_2020} introduced a two-qubit spectroscopy protocol based on continuous-control modulation (spin-locking), to simultaneously reconstruct all self- and cross-spectra. From a theoretical perspective, a framework to classify noise by its fundamental physical nature was developed~\cite{zou_spatially_2023}, distinguishing between classical and quantum noise and two-qubit spectroscopy based on dynamical decoupling was addressed\cite{darrigo_open-loop_2024}.

Machine learning (ML) has recently established itself as a versatile technique
that offers effective diagnostic and control tools for quantum hardware~\cite{MarquardtSPLN2021machine,KrennPRA2023artificial,GebhartNRP2023learning,BrownNJP2021reinforcement,GiannelliPLA2022tutorial,NiunQI2019universal,SgroiPRL2021reinforcement,BanchiNJP2018modelling,TorlaiNP2018neuralnetwork,PalmierinQI2020experimental,barr2024spectral,
barr2025machine, barr2025machine2}. ML-based techniques have been proposed for the
classification of time-correlated noise in qubits, of energy-correlated
fluctuations in qutrits~\cite{MukherjeeMLST2024noise} and networks~\cite{martina_machine_2023}.

In this work, we propose a method for detecting and classifying both temporal and
spatial correlations of noise affecting a small Noise Intermediate Scale Quantum
architecture~\cite{PreskillQ2018quantum} by utilising minimal experimental resources.
A driven two-qubit system is used as a sensor leveraging a control protocol inspired
by Stimulated Raman Adiabatic Passage (STIRAP)~\cite{BergmannRMP1998coherent,VitanovRMP2017stimulated}.
Implementations of this protocol in superconducting devices have been proposed~\cite{siewert_adiabatic_2006,siewert_advanced_2009,falci_design_2013,falci_advances_2017}
and demonstrated~\cite{kumar_stimulated_2016,xu_coherent_2016}. The design we study
for sensing is well-suited to most superconducting-qubit, trapped-ion, and
quantum-dot architectures.

While prior spectroscopic methods~\cite{YonedaNP2023noisecorrelation,von_lupke_two-qubit_2020} provide detailed spectral maps, they require significant experimental overhead and data processing. Similarly, existing ML approaches~\cite{martina_machine_2023,barr2024spectral,barr2025machine,barr2025machine2} rely on measuring time-series data. In contrast, we propose a minimal data acquisition scheme consisting of simple measurements of
STIRAP efficiency with three different combinations of drive amplitudes.
Using synthetic data, we show that the resulting three-component feature vector is
sufficient to train a multi-layer perceptron that reaches accuracies as high as $94\%$
in discriminating both time- and space-correlations of noise. Markovianity of the
quantum dynamics is almost perfectly discriminated, unveiling the ability of the
method to extract information from the imperfections of the protocol.

This approach offers a framework that requires minimal experimental resources for the rapid characterisation of \textit{global} properties of noise affecting quantum hardware. 
Furthermore, in contexts where a full noise estimation is needed, our approach could be a useful preliminary step. Once the nature of the spatial and time correlations of the noise is known, estimation protocols could be more efficiently tailored.

The paper is organized as follows. Section~\ref{sec:Model} introduces the two-qubit
Hamiltonian and the stochastic noise models. In Sec.~\ref{sec:STIRAPin2qubits}, details
of the control protocol and of the performance metrics are given. The machine-learning
framework is described in Section~\ref{sec:ClassificationNN}, the results being presented
and discussed in Section~\ref{sec:Results}. Finally, a summary and an outline of
future perspectives are given in Sec.~\ref{sec:Conclusions}.

\section{\label{sec:Model}The model}
\subsection{The principal system}
\begin{figure}[t!]
    \includegraphics[width=\linewidth]{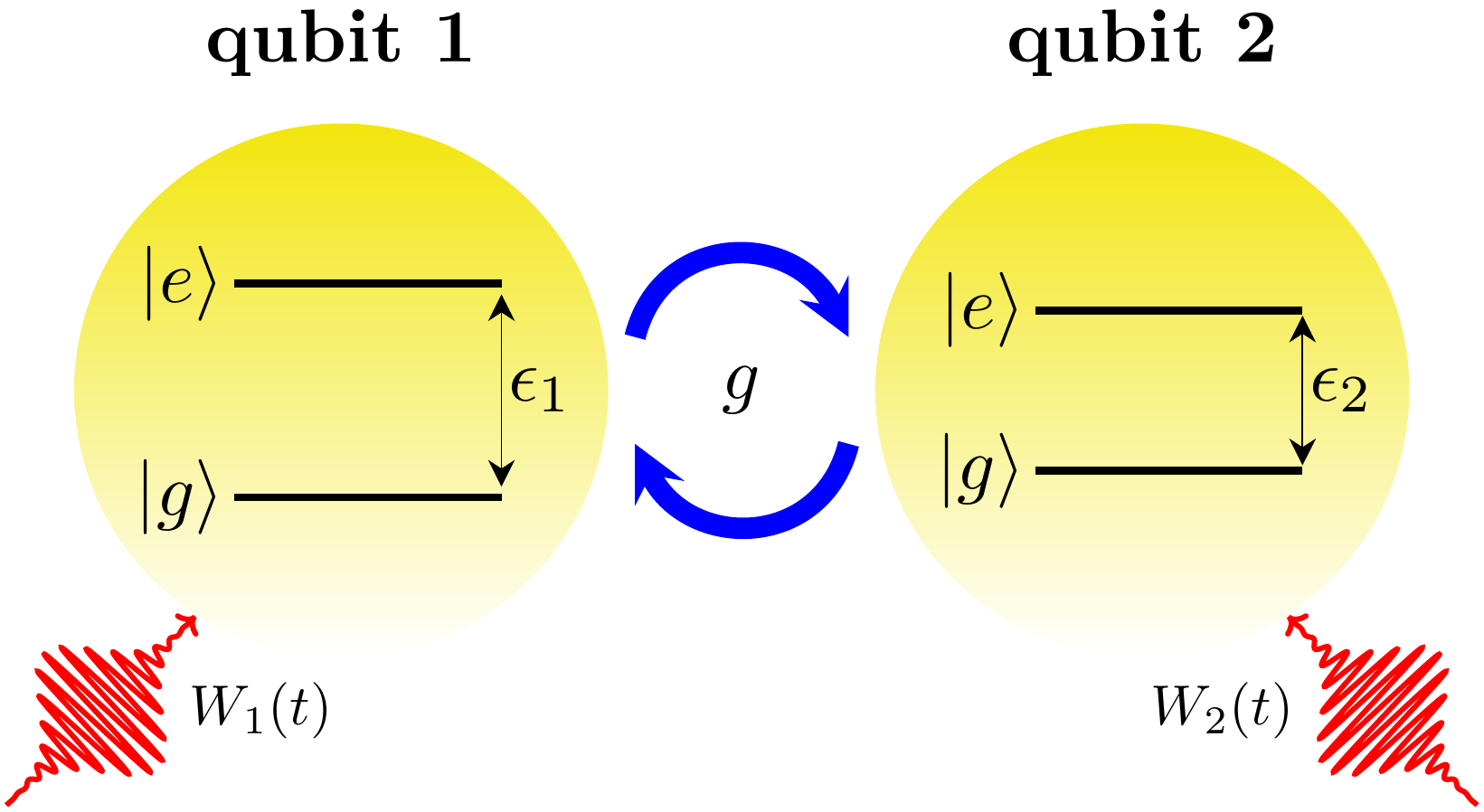}
    \caption{\label{fig:2qubits}Schematic representation of the sensor consisting
    of two qubits, with splittings $\epsilon_{i}$, coupled with strength $g$ and
    driven by external fields $W_{i}(t)$.}
\end{figure}

The sensor is modeled as a two-qubit system with energy splittings
$\epsilon_{i}$, for $i=1,2$, coupled by an Ising-$xx$ interaction of strength
$g$ (see Fig.~\ref{fig:2qubits}). The Hamiltonian reads ($\hbar=1$)
\begin{equation}
    \label{eq:Hamiltonian}H_{S}= -\frac{\epsilon_{1}}{2}\sz[1] -\frac{\epsilon_{2}}{2}
    \sz[2] + \frac{g}{2}\, \sx[1]\sx[2],
\end{equation}
$\sigma^{\alpha}_{i}$ for $\alpha=x,y,z$ being Pauli operators for the $i$-th qubit.

We denote with $\ket{k}$, for $k=0,1,2,3$, the eigenstates of $H_{S}$. Since the
Hamiltonian conserves the parity $\Pi := \sigma_{2}^{z} \sigma_{1}^{z}$ of the
number of excitations $N= (\sigma_{1}^{z}+\sigma_{2}^{z})/2+1$, it admits two invariant
subspaces, namely $\mathscr{H}_{e}:= \mathrm{span}\{\ket{0},\ket{1}\}$ with even
$N$ and $\mathscr{H}_{o}:= \mathrm{span}\{\ket{2},\ket{3}\}$ with odd $N$. The eigenenergies
are respectively $\{-\varepsilon_{e}, \varepsilon_{e}\}$ and $\{-\varepsilon_{o},
\varepsilon_{o}\}$ with
\[
    \varepsilon_{e}= \frac{1}{2}\sqrt{g^{2}+(\epsilon_{1}+\epsilon_{2})^{2}}\quad
    ; \quad \varepsilon_{o}= \frac{1}{2}\sqrt{g^{2}+(\epsilon_{1}-\epsilon_{2})^{2}}
    \;.
\]
The expressions of the eigenvectors are listed in Table~\ref{tab:eigsys}.
\begin{table}[t!]
    \renewcommand{\arraystretch}{1.4}
    \centering
    \begin{tabular}{c|c|l}
        \toprule \tableheadline{Subsp.}             & \tableheadline{Eigenv.} & \tableheadline{Eigenstate}                                                             \\
        \midrule \multirow{2}{*}{$\mathscr{H}_{e}$} & $-\varepsilon_{e}$      & $\ket{0}\equiv-\cos{\frac{\theta_{e}}{2}}\ket{gg}+\sin{\frac{\theta_{e}}{2}}\ket{ee}$  \\
        \cline{2-2} \cline{3-3}                     & $\varepsilon_{e}$       & $\ket{1}\equiv\sin{\frac{\theta_{e}}{2}}\ket{gg}+\cos{\frac{\theta_{e}}{2}}\ket{ee}$   \\
        \hline
        \multirow{2}{*}{$\mathscr{H}_{o}$}          & $-\varepsilon_{o}$      & $\ket{3}\equiv -\cos{\frac{\theta_{o}}{2}}\ket{eg}+\sin{\frac{\theta_{o}}{2}}\ket{ge}$ \\
        \cline{2-2} \cline{3-3}                     & $\varepsilon_{o}$       & $\ket{2}\equiv\sin{\frac{\theta_{o}}{2}}\ket{eg}+\cos{\frac{\theta_{o}}{2}}\ket{ge}$   \\
        \bottomrule
    \end{tabular}
    \caption{\label{tab:eigsys}Eigenvalues and eigenstates of the Hamiltonian~\eqref{eq:Hamiltonian}.
    Here $\{\ket{g},\ket{e}\}$ is the basis of each uncoupled qubit ($g=0$) the mixing
    angles being defined as $\tan{\theta_{e}}= g/(\epsilon_{1}+\epsilon_{2})$ and
    $\tan{\theta_{o}}= g/(\epsilon_{1}-\epsilon_{2})$.}
\end{table}

The system is operated by driving the two qubits locally, the control
Hamiltonian
\begin{equation}
    \label{eq:Hamiltonian_contr_asym}\Hc(t)= \sum_{i}W_{i}(t) \, \sx[i].
\end{equation}
As drive changes the parity, $N \to N \pm 1$, the control Hamiltonian $\Hc(t)$
is ``off-diagonal" in the invariant subspace representation of table \ref{tab:eigsys}
(see App.~\ref{sec:app-model}).

We now introduce a reference scenario where an ideal STIRAP can be operated. For the
sake of simplicity, we take identical qubits $\epsilon_{i}= \epsilon$ so that the odd eigenstates are Bell states
\begin{equation}
    \label{eq:basis_odd}
    \begin{aligned}
         & \ket{2}=\frac{1}{\sqrt{2}}\br{\ket{ge}+\ket{eg}}, \\
         & \ket{3}=\frac{1}{\sqrt{2}}\br{\ket{ge}-\ket{eg}},
    \end{aligned}
\end{equation}
their energy splitting reduces to $g$. Then, if the system
is driven symmetrically, $W_{i}(t) = W(t)$, a selection rule is enforced, implying
that the antisymmetric Bell state $\ket{3}$ is not coupled to the complementary subspace,
$H_{c}\ket{3}=0$. Thus, the driven dynamics is confined to a three-level system.
As the drive doesn't commute with $\Pi$, it implements the so-called ladder
configuration, shown in Fig.~\ref{eq:hamilt_stirap_2qubits}. We anticipate that non-ideal
features emerging when lifting the above assumptions preserve or even improve the
functionality of the noise sensor.
\subsection{The control}
To implement an ideal STIRAP, the control is operated by a two-tone field,
\begin{equation}
    \label{eq:Omega}W(t)={\frac{1}{\beta}}\, \Os(t)\cos{\omega_\mathrm{s} t}+{\frac{1}{\alpha}}
    \,\Op(t)\cos{\omega_\mathrm{p} t},
\end{equation}
where $\omega_{p}=\varepsilon+\frac{g}{2}$ is resonant with the $0-2$ transition
and $\omega_{s}=\varepsilon-\frac{g}{2}$ with the $2-1$ transition, $\Omega_{\mathrm{p/s}}
(t)$ are slowly varying pulse envelopes,
{$\alpha = 2\sin{\br{\tfrac{\theta_e}{2}-\tfrac{\pi}{4}}}$ and $\beta = 2\sin{\br{\tfrac{\theta_e}{2}+\tfrac{\pi}{4}}}$ }.
Here, we use Gaussian $\Omega_{\mathrm{p/s}}(t)$ with width $T$ (see App.~\ref{sec:app-model}).
The full Hamiltonian reads
\begin{equation}
    \label{eq:Hamiltonian_driven_nonoise}
    \begin{aligned}
        H_{S}+\Hc(t) ={} & \varepsilon\br{\ketbra{1}{1} - \ketbra{0}{0}}+\frac{g}{2}\br{\ketbra{2}{2}-\ketbra{3}{3}} \\
        {}               & {}+ W(t)\br{\alpha\ketbra{0}{2}+\beta\ketbra{2}{1}+ \text{h.c.}},
    \end{aligned}
\end{equation}

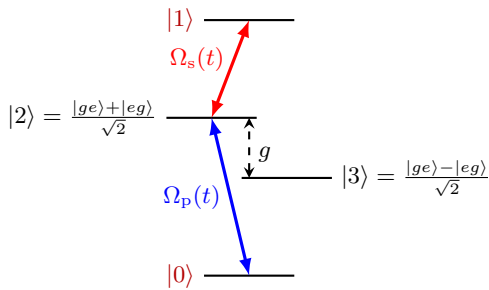
\begin{figure}[t!]
    \begin{tikzpicture}
  \tikzset{>=latex}

  \tikzset{wavy/.style={decorate, decoration = {snake, pre length=3pt,post length=7pt,}}}

  \pgfmathsetmacro{\sL}{1.2} \pgfmathsetmacro{\spacing}{0.50} \pgfmathsetmacro{\Ee}{1.7}
  \pgfmathsetmacro{\g}{0.8} \pgfmathsetmacro{\detp}{0} \pgfmathsetmacro{\det}{0.4}
 




  \pgfmathsetmacro{\Rpos}{0}

  \draw[thick]
    (\Rpos+\spacing,0)
    node[left] {$\ket{0}$}
    --
    +
    +
    (\sL,0);
  \draw[thick]
    (\Rpos,\Ee+\g*0.5)
    node[left] {$\ket{2}= \frac{\ket{ge}+\ket{eg}}{\sqrt{2}}$}
    --
    +
    +
    (\sL,0);
  \draw[thick]
    (\Rpos+2*\spacing,\Ee-\g*0.5) --
    +
    +
    (\sL,0)
    node[right] {$\ket{3}= \frac{\ket{ge}-\ket{eg}}{\sqrt{2}}$};
  \draw[thick]
    (\Rpos+\spacing,2*\Ee)
    node[left] {$\ket{1}$}
    --
    +
    +
    (\sL,0);


  \draw[<->, >=stealth, thick, dashed]
    (\spacing+\sL*0.5,\Ee-\g*0.5) --
    +
    +
    (0,\g)
    node[pos=0.4, right] {$g$};



  \draw[<->, color=blue, very thick]
    (\spacing+\sL*0.5,0) -- (\sL/2,\Ee+\g*0.5)
    node[pos=0.5, left] {$\Op(t)$};

  \draw[<->, color=red, very thick]
    (\sL/2,\Ee+\g*0.5) -- (\spacing + \sL*0.5, 2*\Ee)
    node[pos=0.6, left] {$\Os(t)$};
\end{tikzpicture}
    \caption{\label{fig:system_energyscheme}Energy levels of the two-qubit sensor,
    for identical single-qubit splittings and symmetric driving. The highlighted
    transition implements a ladder driving configuration (see Eq.~\ref{eq:hamilt_stirap_2qubits}).}
\end{figure}

To make explicit contact with STIRAP we introduce a doubly rotating frame and neglect
terms of the drive oscillating at frequencies much larger than $\Omega_{p/s}$.
In the new frame, the control Hamiltonian is approximated as (see App.~\ref{sec:app-model})
\begin{equation}
    \label{eq:hamilt_stirap_2qubits}
    \begin{aligned}
        \tilde{H}_{\rm c}(t) & \approx \left[\frac{\Op(t)}{\sqrt{2}}+\frac{\alpha}{\beta}\frac{\Os(t)}{\sqrt{2}}e^{igt}\right]\ketbra{0}{2}+           \\
                             & +\left[\frac{\Os(t)}{\sqrt{2}}+\frac{\beta}{\alpha}\frac{\Op(t)}{\sqrt{2}}e^{-igt}\right]\ketbra{1}{2}+ \text{h.c.}\; .
    \end{aligned}
\end{equation}
The standard Rotating Wave Approximation (RWA) is obtained by eliminating the
terms that oscillate with frequency $g$, which would be accurate when $g \gg \Omega
_{p/s}$. In this limit, the control Eq.\ref{eq:hamilt_stirap_2qubits} implements
the usual 3-level ``ladder" configuration (see Fig.~\ref{fig:system_energyscheme})
admitting as an eigenvector a ``dark state", i.e. a coherent superposition of
$\ket{0}$ and $\ket{1}$. Its adiabatic dynamics (see Eq.~(\ref{eq:globaladiabaticitycondition}))
under a ``counterintuitive" pulse sequence, where $\Omega_{p}(t)$ is delayed by
a time $\tau$ with respect to $\Omega_{s}(t)$ (see Eq.~\eqref{eq:Gaussian_pulses}),
implements the ideal 3-level STIRAP protocol (see App.~\ref{sec:STIRAP}), which results
in coherent population transfer $\ket{0}\approx \ket{gg}\to \ket{1}\approx \ket{ee}$.
During ideal STIRAP, the intermediate state $\ket{2}$ is never populated due to destructive
interference. This scenario is also substantially valid under the action of the
{\em full Hamiltonian} Eq~\eqref{eq:Hamiltonian_driven_nonoise}, as shown in Fig.~\ref{fig:stirap_2qubits}.

\subsection{\label{sec:Noise}The Noise}
\begin{figure}[t!]
    \includegraphics{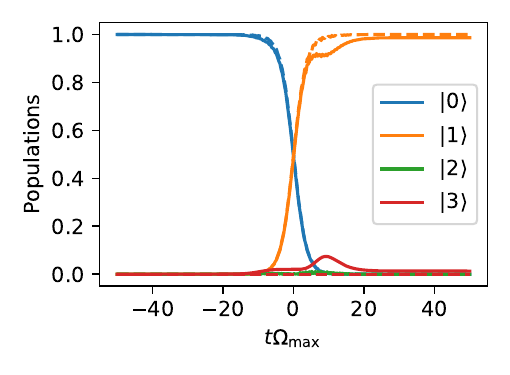}
    \caption{\label{fig:stirap_2qubits} Population histories of a system of two
    ultrastrongly coupled qubits, $g=0.6\,\epsilon$. The dashed lines represent
    the dynamics with the full Hamiltonian Eq.~(\ref{eq:Hamiltonian_driven_nonoise}).
    Pulses given by Eq.~\eqref{eq:Gaussian_pulses} with parameters
    $\Opsmax=\Omega_{\mathrm{max}}=0.05\,\epsilon$, $\epsilon T =2000$ and $\tau
    =0.7 T$. The evolution is close to ideal STIRAP. Quasistatic noise introduces
    imperfections (solid lines). Here $\delta_{1}=0.002 \, \epsilon$ and
    $\delta_{2}=-0.002\, \epsilon$.}
\end{figure}

We consider local noise affecting the level splittings of each qubit which is modeled
by the Hamiltonian
\begin{equation}
    \label{eq:Hamiltonian_noise}\Hnoise(t) = - \frac{1}{2}\big[\delta_{1}(t)\sz[1
    ] + \delta_{2}(t)\sz[2] \big].
\end{equation}
where $\delta_{i}(t)$ for $i=1,2$ are classical stochastic processes~\cite{Mandel1995optical}.
This model accounts for the main mechanism degrading coherence during the
adiabatic passage phase of population transfer~\cite{BergmannRMP1998coherent,VitanovRMP2017stimulated},
which is the heart of the quantum sensing protocol. Physically, it may arise in many
cases, in particular by such as charge noise~\cite{YonedaNP2023noisecorrelation,ZorinPRB1996background,PaladinoRMP2014}. Expressing$\Hnoise(t)$in the eigenbasis of$H_S$we obtain
\begin{equation}
    \label{eq:Hamilt_noise}
    \begin{aligned}
        \Hnoise(t) & =\frac{1}{2}\big(\delta_{1}+\delta_{2}\big) \cos{\theta_e}\big( \ketbra{1}{1}-\ketbra{0}{0}\big)+{}                                               \\
        +          & \frac{1}{2}\Big[ \big(\delta_{1}-\delta_{2}\big) \ketbra{3}{2}-\frac{g }{2 \varepsilon}(\delta_{1}+\delta_{2}) \ketbra{0}{1}+ \textrm{h.c.}\Big].
    \end{aligned}
\end{equation}
It is seen that noise induces nonideal features for STIRAP: (i) stochastic fluctuations to the energy splitting in the even subspace $\mathscr{H}_e$, which is the
``trapped" subspace for STIRAP, and produce substantial dephasing; (ii) a coupling
g between states $\ket{2}-\ket{3}$ inducing leakage from the two-level subspace; (i
ii) a coupling between$\ket{0}-\ket{1}$. All these features are still compatible
with STIRAP but produce a slightly different output (see Fig.~\ref{fig:stirap_2qubits}
, solid lines) providing information on the noise.
  For the data-analysis we consider six classes of noise, according to their time
and space correlations:
\begin{itemize}
    \item {\bf Time-correlated (non-Markovian) noise:} In this context we address
        specifically the case of \textit{quasistatic noise}, where the correlation
        time is larger than the duration of a single run of the protocol. Noise is reduced to random variables $\delta_{i}(t)$
        constant during a single run, but varying for different runs. This is
        the leading effect of $1/f$ noise~\cite{paladino2002,falci2005,PaladinoRMP2014,thorwart2025}
        in solid-state quantum architectures. The values of $\delta_{i}(t)$ are picked
        from a Gaussian distribution with zero mean and {standard deviation $\sigma=10^{-2}\epsilon = 0.2\sqrt{\left({\Opmax}^{2}+{\Osmax}^{2}\right)/2}$},
        and describe the cumulative effects of many independent microscopic
        sources which are weakly individually coupled~\cite{falci2005}. Concerning
        space correlations we focus on three distinct classes:
        \begin{enumerate}[label=(\arabic{*})]
            \item Correlated: $\delta_{2}(t)= \eta \delta_{1}(t)$ with $\eta>0$;

            \item Anti-correlated: $\delta_{2}(t) = \eta \delta_{1}(t)$ with
                $\eta<0$;

            \item Uncorrelated: independent $\delta_{2}(t)$ and $\delta_{1}(t)$.
        \end{enumerate}

    \item {\bf Markovian noise:} We consider zero-mean $\avg{\delta_i(t)}=0$,
        delta-correlated
        $\avg{\delta_i(t)\delta_i(t^\prime)}= \gamma\delta(t-t^{\prime})$
        stochastic processes, thus the dynamics of the two-qubits is described by
        a Markovian map and does not depend on its history. Here we consider
        again three instances of space-correlations:
        \begin{enumerate}[label=(\arabic{*}), resume]
            \item Correlated: $\delta_{2}(t)= \eta \delta_{1}(t)$, with $\eta >0$;

            \item Anti-correlated: $\delta_{2}(t) = \eta \delta_{1}(t)$, with $\eta
                <0$.

            \item Uncorrelated: independent $\delta_{2}(t)$ and $\delta_{1}(t)$.
        \end{enumerate}
\end{itemize}
 While real hardware noise may exhibit more complex spectral features, non-Gaussian statistics, or additional couplings, the aim of the present work is not to reproduce a specific experimental device. Rather, we focus on demonstrating that global properties of noise correlations-both temporal and spatial-can be reliably discriminated using minimal experimental information. The proposed framework is general and can be straightforwardly extended to incorporate other types or platform-specific noise models.
\section{\label{sec:STIRAPin2qubits}Population transfer in the noise sensor}
Our main goal is to classify the different noises by Machine Learning, using the
efficiency of a STIRAP-like protocol, as a fingerprint. The full Hamiltonian des
cribing the driven dynamics of the sensor in the presence of noise is given by $H_S+H_c(t)+\Hnoise(t)$.
The main effect of noise (as described in Sec.~\ref{sec:Model}) is due to the dia
gonal entries of $\Hnoise(t)$in the eigenbasis of $H_S$, see eq.~(\ref{eq:Hamilt_noise}
), since the dark state is not anymore an eigenstate of the full (dressed) Hamiltonian. As a consequence, adiabatic patterns do not lead to population transfer. However, for small enough $\delta_i \ll \Opsmax$ almost complete population transfer
occurs via diabatic transitions.  In addition, noise may also couple the antisym
metric Bell state $\ket{3}$ with $\ket{2}$, the dynamics now involving the whole 4-level Hilbert space. Remarkably, $\ket{3}$ gets populated during time evolution even
if the noise is small $\delta_{1},\delta_{2}\ll \varepsilon$. An example of population histories in the four-level system is shown in Fig.~\ref{fig:stirap_2qubits}
(solid lines).
\subsection{\label{sec:efficiency}Efficiency}
\begin{figure}[t!]
    \includegraphics{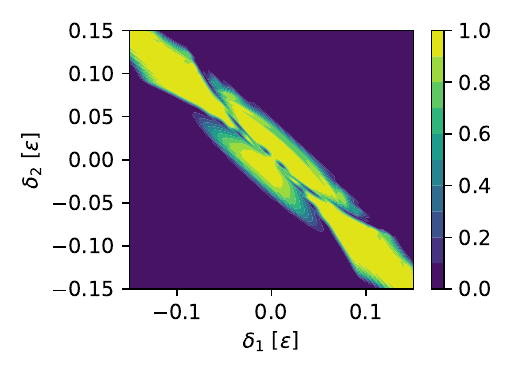}
    \caption{\label{fig:stabplot_noise}Efficiency of the STIRAP-like protocol versus
    the quasistatic noises $\delta_{1}$ and $\delta_{2}$ for equal peak amplitudes
    $\Opsmax$. The parameters are the same as in Fig.~\ref{fig:stirap_2qubits}.}
\end{figure}

In the foregoing we will study the efficiency of the protocol defined as the final population of the state $\ket{ee}$, that is
\begin{equation}
    \label{eq:efficiency}\xi=\lim_{N\to\infty}\frac{1}{N}\sum_{r=1}^{N}\xi^{(r)}
\end{equation}
where $\xii=\matrixel{ee}{\rhof^{(r)}}{ee}$ and $\rhof^{(r)}$ is the density matrix of the system at the final time $\tf$ for the $r$-th noise realization. This is not the usual efficiency of STIRAP given by the final population of the target state $\ket{1}$ which is entangled, measurements in the local basis providing a simpler data-acquisition scheme.  For quasistatic noise, which remains constant during the individual trajectory, the index $r$ can be uniquely associated with the two real values $\delta_{1}^\prime{(r)}$ and $\delta_{2}^\prime{(r)}$, which are specific values of the random variables $\delta_{1}$ and $\delta_{2}$, respectively. Therefore we write $\xii= \xi(\delta_{1}^{(r)},\delta_{2}^{(r)})$ and express the efficiency Eq.~\eqref{eq:efficiency}via the joint probability distribution $p(\delta_{1},\delta_{2})$\begin{equation}
    \label{eq:efficiency_quasistatic_uncorr}\xi=\int \de\delta_{1}\, d\delta_{2}\,
    \xi(\delta_{1},\delta_{2})\,p(\delta_{1},\delta_{2}).
\end{equation}
The efficiency for equal peak values of the drives, $\Osmax=\Opmax$, is reported in Fig.~\ref{fig:stabplot_noise} as a function of the values of the quasistatic noises $\delta_{i}$. We emphasize that our goal is not to achieve efficient population transfer, but rather to leverage the protocol's sensitivity to extract information about noise, which reduces the efficiency. Fig.~\ref{fig:stabplot_noise}
suggests that the average efficiency strongly depends on the correlations of the
random variables $\delta_i$. This information will be accessed by considering three different ratios $\Osmax/\Opmax$ of the peak pulse amplitudes (see Fig.~\ref{fig:stabplot_noise_kvar}).
 For perfectly correlated or anti-correlated noise,$\delta_{2}^{(r)}=\eta\delta_{1}^{(r)}$,
Eq.~(\ref{eq:efficiency_quasistatic_uncorr}) simplifies
\begin{equation}
    \label{eq:efficiency_quasistatic_corr}\xi=\int d\delta_{1}\, \xi(\delta_{1},\eta
    \delta_{1})\,p(\delta_{1}).
\end{equation}
For uncorrelated noises the joint probability factorizes.  For Markovian noise,
the efficiency is obtained by solving a Lindblad master equation (see Appendix.~
\ref{sec:LME})
\begin{equation}
    \label{eq:lme_equation}
    \begin{aligned}
        \avg{\dot{{\rho}}(t)}= & -i\left[H(t),\rho(t)\right] +                                                                                 \\
                               & + \sum_{k}\gamma_{k} \br{{O^\dagger_k}{\rho}(t){O_k}-\frac{1}{2} \left[{O^\dagger_k O_k},{\rho}(t)\right]_+},
    \end{aligned}
\end{equation}
where$\avg{.}$denotes the average over the stochastic process. Uncorrelated noise
is described by two jump operators which in the eigenbasis of $H_S$ read
\begin{equation}
    \label{eq:lme-uncorr}{ \begin{aligned}O_{1} =&\frac{1}{2}\Big[\cos{\theta_e}\big(\ketbra{1}{1}- \ketbra{0}{0}\big) + \big( \ketbra{2}{3}+\ketbra{3}{2}\big) + \\&\frac{g}{2\varepsilon}\ketbra{0}{1}+ h.c\Big]\\ O_{2} =&\frac{1}{2}\Big[\cos{\theta_e}\big(\ketbra{1}{1}- \ketbra{0}{0}\big) - \big( \ketbra{2}{3}+\ketbra{3}{2}\big) + \\&\frac{g}{2\varepsilon}\ketbra{0}{1}+ h.c\Big]\\\end{aligned}}
\end{equation}
A single jump operator describes instead correlated/anticorrelated noise
\begin{equation}
    \begin{aligned}
        O = & {\frac{1}{2}\Bigg[\br{\frac{1+\eta}{\sqrt{\abs{\eta}}}\cos{\theta_e}}\br{\ketbra{1}{1}-\ketbra{0}{0}}+} \\
            & {-\br{\frac{g}{2\varepsilon}\frac{1+\eta}{\sqrt{\abs{\eta}}}}\br{\ketbra{0}{1}+\ketbra{1}{0}}}          \\
            & +\br{\frac{1-\eta}{\sqrt{\abs{\eta}}}}\br{\ketbra{2}{3}+\ketbra{3}{2}}\Bigg].
    \end{aligned}
\end{equation}
\section{\label{sec:ClassificationNN}Machine Learning model and Data generation}
\subsection{\label{sec:MLmodel}ML model}
We employ a neural network (NN), namely a Multi Layer Perceptron (MLP), to
perform supervised learning \cite{Goodfellow2016deep, bishop2006pattern} and
classify the type of noise affecting the two-qubit system. A N Nis a parameterized function capable of approximating a broad class of
target functions~\cite{HORNIK1989359}; the specific architecture we use is reported in Table~\ref{tab:NNstructure}.
\begin{table}[!ht]
    \renewcommand{\arraystretch}{1.4}
    \centering
    \begin{tabular}{c|c|c}
        \toprule \tableheadline{Layer} & \tableheadline{$\#$ Neurons} & \tableheadline{Activation Function} \\
        \midrule Input                 & $3$                          &                                     \\
        \hline
        Hidden 1                       & $64$                         & ReLU                                \\
        \hline
        Hidden 2                       & $32$                         & ReLU                           \\
        \hline
        Hidden 3                       & $32$                         & LeakyReLU                           \\
        \hline
        Hidden 3                       & $32$                         & LeakyReLU                           \\
        \hline
        Output                         & $6$                          & Softmax                             \\
        \bottomrule
    \end{tabular}
    \caption{\label{tab:NNstructure}Layout of the neural network for classifying
    the types of noise.}
\end{table}

Supervised learning consists in fitting a function approximator (in this case,
the NN) that maps inputs to outputs by minimizing a cost function over a labeled
dataset. This procedure is referred to as \textit{training} the neural network~\cite{MarquardtSPLN2021machine,Goodfellow2016deep}.
The dataset consists of input-output pairs
$(\vect{x}_{i},\hat{\vect{y}}_{i})_{i= 1,2,\dots,N}$, where each input $\vect{x}_{i}$
is associated with a corresponding output $\hat{\vect{y}}_{i}$, also called
label. Here the label $\hat{\vect{y}}_{i}$ for each sample $i$ is an integer $a=0
,1,2,3,4, 5$ representing one of the noise classes described in Sec.~\ref{sec:Noise}.
The output $\vect{y}_{i}$ of the NN is a $6$ dimensional real vector whose components
$y_{i,j}$ represent the probabilities that the noise affecting the system of sample
$i$ is of class $j$. The use of the \textit{Softmax} \cite{Goodfellow2016deep, Geron2023handson}
activation in the output layer (see Tab.~\ref{tab:NNstructure}) ensures that the
model outputs form a normalized probability distribution. We minimize the
\textit{sparse categorical cross-entropy}\cite{glorot2011deep} cost function
\begin{equation}
    \label{eq:sparsecategoricalcrossentropy}C = -\frac{1}{N}\sum_{i=1}^{N}\log \br
    {y_{i,\hat{y}_i}},
\end{equation}
which is well-suited for multi-class classification tasks.

This choice encourages the model to assign high probability to the correct label. The
training process is monitored through the classification accuracy, defined as
\begin{equation}
    \label{eq:accuracy}A = \frac{1}{N}\sum_{i=1}^{N}\boldsymbol{\delta}\big( \argmax
    _{j}y_{ij},\ \hat{\vect{y}}_{i}\big),
\end{equation}
where $\boldsymbol{\delta}$ is the Kronecker delta, and $\argmax_{j}$ returns
the index $j$ of the largest component of the corresponding vector.

More details on the neural network architecture and training procedure are
provided in the Appendix~\ref{sec:NN}.

\subsection{\label{sec:datageneration}Data generation}
We use as input to the ML model the STIRAP efficiencies measured under three
different driving conditions~\cite{MukherjeeMLST2024noise}
\begin{enumerate}[label=(\roman{*})]
    \item $\Opmax = \Osmax$,

    \item $\Opmax = 2\Osmax$,

    \item $\Opmax = \Osmax/2$,
\end{enumerate}
with the constraint $(\Opmax)^{2}+ (\Osmax)^{2}$ constant. This approach is advantageous
as it provides high sensitivity to noise while requiring low experimental effort
and is straightforward to implement in most physical systems. In fact the
measurement has to be performed only at the end of (and not during) the
evolution. This eliminates the necessity for time series data and may drastically
reduce the duration of the experiment. In fact, by relying exclusively on final-state populations, the protocol avoids time-resolved measurements or continuous monitoring. This substantially reduces experimental overhead, data acquisition and storage requirements, and makes the approach particularly suitable for scalable implementations.

To generate the dataset, we numerically simulate the dynamics of the two-qubits system
and evaluate the efficiency as reported in Sec.~\ref{sec:efficiency}. Each input
data point is then a three dimensional Real vector $\vect{x}= (\xi_{\Op=\Os}, \xi
_{\Op>\Os}, \xi_{\Op<\Os})$.

For correlated (anti-correlated) non-Markovian noise we generate each data sample
by randomly selecting the correlation parameter $\eta$ within the range $[0.1 , 5
]$ ($[-5, -0.1]$). For each choice of $\eta$, we numerically evaluate the efficiency,
eq.~\eqref{eq:efficiency_quasistatic_corr}, under the three pulse conditions outlined
earlier, yielding a single sample.

For uncorrelated non-Markovian noise instead, as the two random variables $\delta
_{1}$ and $\delta_{2}$ are independent, we vary the standard deviation of the
Gaussian distributions $p_{1}(\delta_{1})$ and $p_{2}(\delta_{2})$ respectively
within the range $\sigma_{1}, \sigma_{2}\in [\frac{\sigma}{5}, 5\sigma ]$, with $\sigma
=10^{-2}\epsilon$ before calculating every data sample. We then numerically
calculate the efficiencies under the three pulse condition using eq.~\eqref{eq:efficiency_quasistatic_uncorr}.

In the context of Markovian, correlated (anti-correlated) noise, the
efficiencies are calculated numerically solving the Lindblad master equation, eq.~\ref{eq:lme_equation}.
In this particular instance, not only does $\eta$ fluctuate within the specified
ranges, like correlated (anti-correlated) non-Markovian scenarios from one
sample to another, but also we randomly select the decay parameter $\gamma \in [1
0^{-4}, 10^{-3}]\epsilon$. Whereas, for uncorrelated Markovian noise we randomly
select the decay parameters $\gamma_{1}, \gamma_{2} \in [10^{-4}, 10^{-3}]\epsilon$.
For each noise class, we generate $5 00$ samples. The total dataset is split
into a $3:1:1$ ratio for training, validation, and test set respectively.

 \section{\label{sec:Results}Results}
\begin{figure}[t!]
    \includegraphics[width=\linewidth]{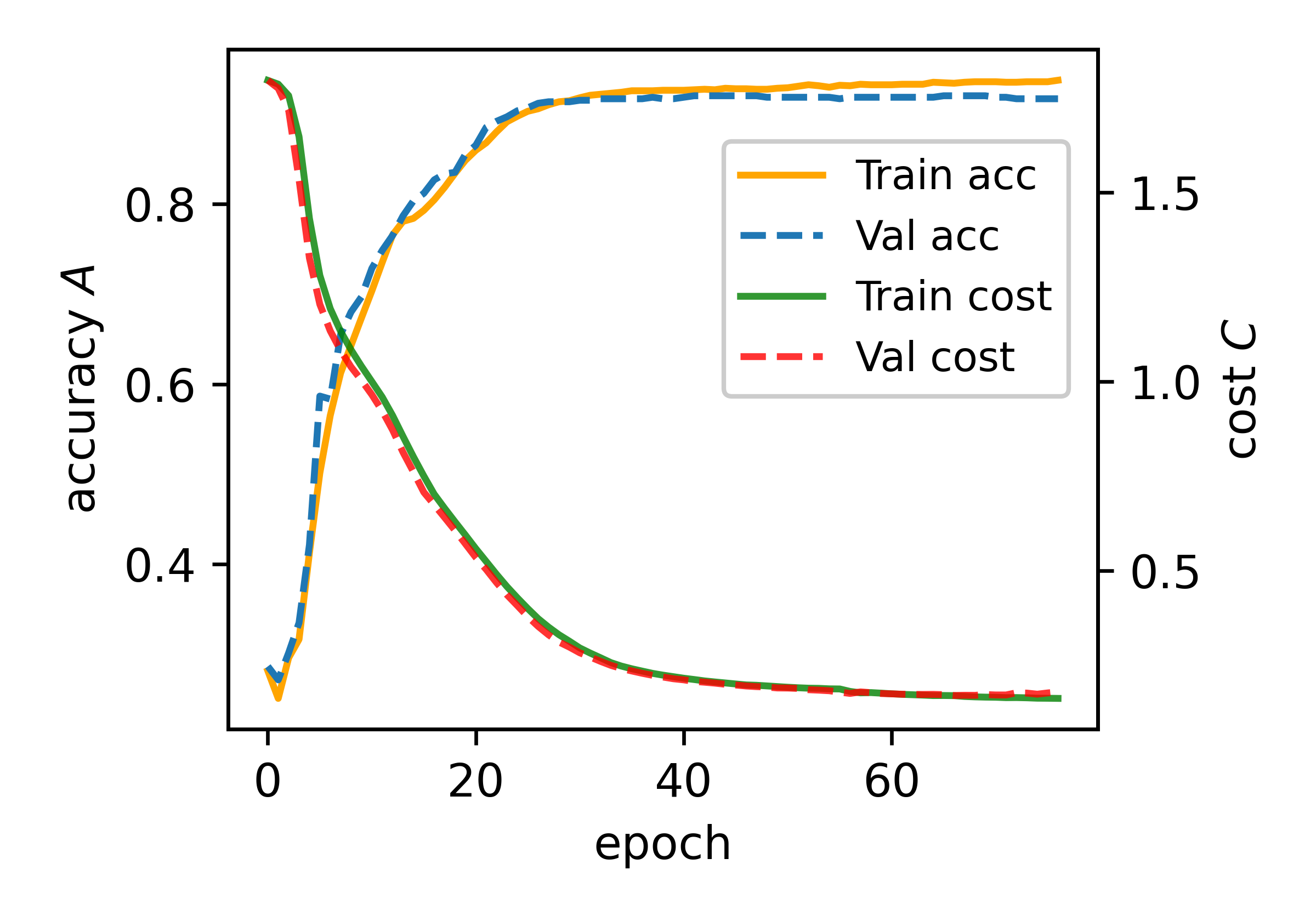}
    \caption{\label{fig:training} Accuracy~\eqref{eq:accuracy} and value of the cost function,
    Eq.~\eqref{eq:sparsecategoricalcrossentropy}, for the training (solid
    lines) and validation (dashed lines) sets versus the number of epochs of
    training. The accuracy on the test set is $A \approx 94\%$.}
\end{figure}

\begin{figure}[t!]
    \includegraphics[width=0.9 \linewidth]{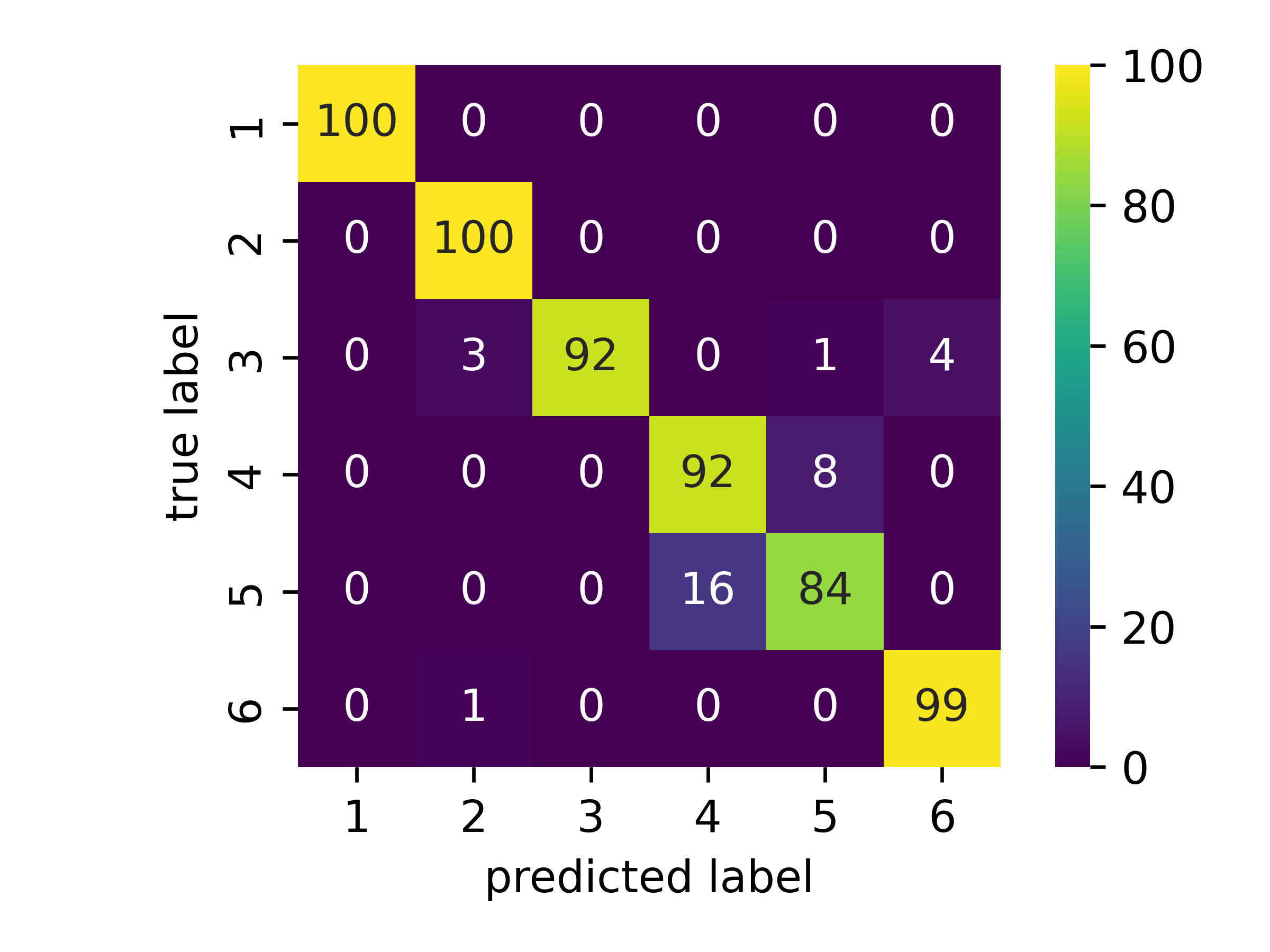}
    \caption{\label{fig:confusion_matrix}. Confusion matrix of the MLP model for
    classifying the noise types described in Sec.~\ref{sec:Noise}. Each row represents
    the true noise class, while each column corresponds to the predicted class. The
    classification reaches an accuracy of $\approx 94\%$}.
\end{figure}

We train the NN model described in Tab.~\ref{tab:NNstructure}. The output
$\vect{y}_{i}$ of the neural network for a given input vector $\vect{x}_{i}$, denoted
as is compared to the corresponding true label $\hat{\vect {y}}_{i}$. The
training progress of the model is illustrated in Fig.~\ref{fig:training}, where
panel (a) shows the classification accuracy $A$, eq.~\ref{eq:accuracy}. The accuracy
increases with the number of training epochs. Panel (b) shows that the cost
function $C$, eq.~\ref{eq:sparsecategoricalcrossentropy}, decreases accordingly.
After about $70$ training epochs, the model achieves a classification accuracy of $(94
\pm2)\%$, the variation being due to the random initialization of model
parameters, as well as the stochastic shuffling and splitting of the dataset.
The classification performance of the model is further analyzed using the
confusion matrix in Fig.~\ref{fig:confusion_matrix}, which compares the true
noise classes with the predicted ones. Each row corresponds to a true noise class,
while each column represents the predicted class. The six diagonal elements indicate
the percentage of correct predictions for each noise class.
The model achieves an accuracy of $99\%$ in distinguishing between non-Markovian
and Markovian noise. It correctly classifies space correlations of non-Markovian
noise with an accuracy of $97.3\%$, whereas within the Markovian class, the classification
accuracy is $91.7\%$.

\begin{figure}[t!]
    \includegraphics{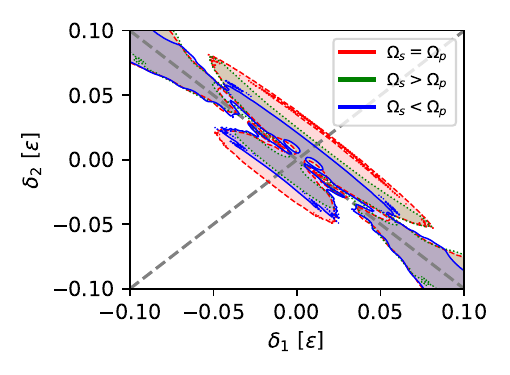}
    \caption{\label{fig:stabplot_noise_kvar}Contour plot of the efficiency of the
    STIRAP-like protocol. The lines enclosing the different regions correspond
    to $75 \%$ efficiency of population transfer for different ratios between
    the peak values of the pulse amplitudes. The parameters are the same of fig.~\ref{fig:stabplot_noise}.
    The dashed lines $\delta_{1}=\pm\delta_{2}$ emphasize the different statistics
    of correlated and anticorrelated quasistatic noise.}
\end{figure}

Insight on how non-Markovian noise is classified is provided by Fig.~\ref{fig:stabplot_noise_kvar}
where the efficiency vs the realizations $(\delta_{1},\delta_{2})$ of quasistatic
noise is reported for different ratios $\Osmax/\Opmax$ of the peak amplitudes of
the drives. Non-Markovian quasistatic noises randomly shift the energy splitting
of the corresponding qubit at each repetition of the protocol. Consequently, for
fully correlated (anticorrelated) quasistatic noise, the overall efficiency is obtained
as the weighted integral of the single trajectory efficiency over a line $\delta_{2}
= \eta \delta_{1}$, see eq.~\eqref{eq:efficiency_quasistatic_corr}. Since the efficiency
profiles depend on the driving conditions and are different for $\eta>0$ or $\eta
<0$, also the average efficiencies are expected to depend on driving conditions
and correlations. Even if it is not immediate that this may lead to
efficient discrimination, data analysis by ML provides a robust solution to this
problem.

For uncorrelated quasistatic noise, where the weighted integral is over the
whole $\delta_{1}- \delta_{2}$ plane (see eq.~\eqref{eq:efficiency_quasistatic_uncorr}),
the situation is less clear. For Markovian noise the efficiency for each
trajectory cannot be represented as a single point $(\delta_{1},\delta_{2})$ and
the analysis is even more involved. Yet discrimination by ML is to a large extent successful.

We notice that, compared to an earlier work on correlated noise in three-level
systems~\cite{MukherjeeMLST2024noise}, where ML was unable to detect energy correlations
between different Markovian noises, the two-qubit setup allows to discriminate space correlations of Markovian noise. This enhanced classification ability stems
from the noise-induced mixing of the intermediate states $\ket{2}$ - $\ket{3}$ and
$\ket{0}$ - $\ket{1}$ (see Eq.~\eqref{eq:Hamilt_noise}). As a consequence, the dynamics
involves the whole 4-level Hilbert space to an extent {\em depending on the correlations}
which makes the discrimination effective also for Markovian noise.

\begin{figure}[!ht]
        \includegraphics{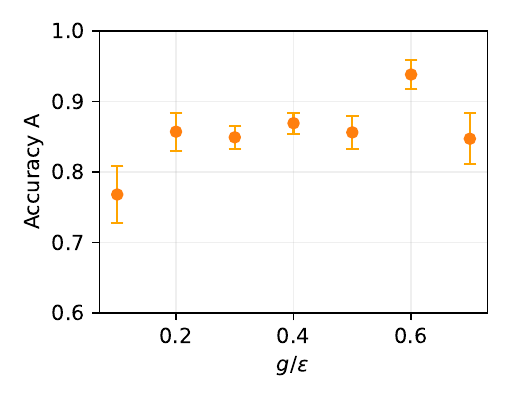}
        \caption{\label{fig:acc_vs_g} Accuracy $A$ vs the qubit-qubit coupling strength $g$. Other parameters are reported in Sec.~\ref{sec:datageneration}. The error is estimated as the semi-dispersion over 10 independent training runs per dataset, with randomized parameter initialization and data splitting.}
    \end{figure}

    Figure~\ref{fig:acc_vs_g} reports the noise classification accuracy $A$, eq.~\eqref{eq:accuracy} as a function of the qubit-qubit coupling strength $g$. For each value of $g$ we generated a labeled dataset of $3000$ samples, tuned the neural network architecture, and trained the model to evaluate the corresponding classification accuracy.
    For $g\geq 0.2\epsilon$ , the accuracy always remains higher than $80\%$, demonstrating the robustness of the protocol versus different values of $g$. The maximum accuracy obtained is $A\approx 0.94$ and is reached at $g=0.6\epsilon$.

\section{\label{sec:Conclusions}Conclusions}
In this work, we have proposed and numerically validated a machine learning-assisted quantum sensing which categorizes classical noises affecting a two-qubit system.
The protocol extracts global information on time and/or space correlations of
noise. We defined and successfully classified six distinct noise classes, namely
correlated/anti-correlated/uncorrelated non-Markovian and Markovian noises. Our approach
leverages the sensitivity of coherent population transfer by a non-ideal STIRAP-like
protocol to the amplitudes of the drives. Variations in the efficiency provide a
fingerprint of the underlying noise correlations.

By measuring only the final population of the doubly excited state $\ket{ee}$
under the three distinct driving conditions, the protocol achieves a classification
accuracy higher that $80\%$ for several values of the qubit-qubit coupling strength $g$, with a peak of $A\approx 94\%$ at $g=0.6\epsilon$. Furthermore, discrimination of time-correlations
is nearly perfect. Remarkably, this approach requires only three inputs to a
shallow neural network, with no need for time-series data or real-time monitoring and requires a simple driving scheme, thus having a minimal experimental
overhead. This approach of noise classification and characterization differs from standard methods such as reconstruction of noise power spectra or full process tomography. 
While these techniques provide more detailed information about the environment, they require much more time and experimental resources. In contrast, our method quickly extracts global properties of the noise using only a small number of measurements.

Several driving control schemes have been found\cite{BergmannRMP1998coherent,VitanovRMP2017stimulated,di_stefano_population_2015,di_stefano_coherent_2016,brown_reinforcement_2021}
which can be employed in different implementations of the sensor. Data can in principle be collected by averaging single-shot measurements of each qubit. An experimentally
less demanding procedure is a continuous {measurement~\cite{di_stefano_nonequilibrium_2018}}
of the decay of each atom into a transmission line. A large signal is obtained
if such decay is faster than the individual atomic non-radiative decay~\cite{giannelli_detecting_2024}.
In the simplest instance, the transmission line is coupled to the mode during
the whole protocol. As analized in {Ref.~\onlinecite{giannelli_integrated_2023}},
despite the apparent invasivity, the measurement backaction is expected to have little
effect on STIRAP, since the main effect is the decay of the final state when
populated, which is the measurement itself.

Differently from the previous work on correlated noise in three-level systems~\cite{MukherjeeMLST2024noise},
the two-qubit setup allows to discriminate space correlations of Markovian noise.
This enhanced sensitivity arises from the richer physics of the four-level system,
where noise may induce deviations from the ideal protocol. These latter provide qualitatively
new data which improve the discrimination power of the supervised learning analysis.

The main goal of this work is to demonstrate the feasibility of detecting noise correlations using a small set of features obtained from an advanced quantum control protocol, and analyzed by supervised learning.
Future work will broaden the scope of our approach to achieve a finer
classification of correlations by exploiting other non-idealities of the
protocol, by utilizing alternative control schemes, and by 
identifying additional input features. Simultaneous optimization of parameters of the sensor besides 
the coupling $g$, such as the STIRAP parameters $\Opsmax$ and $\tau$, could further improve the accuracy. Furthermore, the size of the sensor can be increased by considering array of qubits which may provide richer noise fingerprints: collective phenomena in these multi-qubits systems are also expected to provide an advantage in achieve a finer classification. Another
important step would be tailoring unsupervised learning strategies for noise
classification. We believe that the synergy between coherent control schemes and
ML approaches, by further enhancing classification sensitivity and broadening
applicability to diverse quantum platforms, will play a key role in the next-generation
quantum-hardware diagnostics.

\appendix

\section{\label{sec:app-model}Control Hamiltonian}
The control Hamiltonian Eq.(\ref{eq:Hamiltonian_contr_asym}) describes local
drives acting on each qubit. In the basis of the eigenstates of $H_{S}$ it reads
\begin{equation}
    \label{eq:Hamiltonian_contr_full_asymmetric}
    \begin{aligned}
        \Hc(t) = & -\brr{W_1\sin{\br{\tfrac{\theta_o-\theta_e}{2}}}+W_2\cos{\br{\tfrac{\theta_o+\theta_e}{2}}}}\ketbra{0}{2}+    \\
        +        & \brr{W_1\cos{\br{\tfrac{\theta_o-\theta_e}{2}}}+W_2\sin{\br{\tfrac{\theta_o+\theta_e}{2}}}}\ketbra{1}{2}+     \\
        +        & \brr{W_1\cos{\br{\tfrac{\theta_o-\theta_e}{2}}}-W_2\sin{\br{\tfrac{\theta_o+\theta_e}{2}}}}\ketbra{0}{3}+     \\
        +        & \brr{W_1\sin{\br{\tfrac{\theta_o-\theta_e}{2}}}-W_2\cos{\br{\tfrac{\theta_o+\theta_e}{2}}}}\ketbra{1}{3}+h.c.
    \end{aligned}
\end{equation}
resulting ``off-diagonal" in the invariant subspace representation (table \ref{tab:eigsys})
since operators $\sigma_{i}^{x}$ in the drive change the parity. From Eq.~\ref{eq:Hamiltonian_contr_full_asymmetric},
we identify under which either state $\ket{2}$ or $\ket{3}$ are decoupled in the
dynamics. This happens when $\theta_{o}={\pi}/{2}$ and $\tfrac{W_1}{W_2}=\pm1$, where
the positive sign applies to state $\ket{3}$, and the negative sign applies to
$\ket{2}$.

For symmetric driving, $W_{1}=W_{2}=W$, the above equation simplifies
\begin{equation}
    \label{eq:Hamiltonian_contr_subspace}
    \begin{aligned}
        \Hc(t) & = 2 W(t)\left[\sin{\br{\tfrac{\theta_o}{2}+\tfrac{\pi}{4}}}\sin{\br{\tfrac{\theta_e}{2}-\tfrac{\pi}{4}}}\ketbra{0}{2}\right.+ \\
        +      & \sin{\br{\tfrac{\theta_o}{2}+\tfrac{\pi}{4}}}\sin{\br{\tfrac{\theta_e}{2}+\tfrac{\pi}{4}}}\ketbra{1}{2}+                      \\
        +      & \sin{\br{\tfrac{\theta_o}{2}-\tfrac{\pi}{4}}}\sin{\br{\tfrac{\theta_e}{2}-\tfrac{\pi}{4}}}\ketbra{0}{3}+                      \\
        +      & \left.\sin{\br{\tfrac{\theta_o}{2}-\tfrac{\pi}{4}}}\sin{\br{\tfrac{\theta_e}{2}+\tfrac{\pi}{4}}}\ketbra{1}{3}+h.c.\right],
    \end{aligned}
\end{equation}
and condition $\theta_{o}= \tfrac{\pi}{2}$ is found, corresponding to identical qubits,
$\epsilon_{i}=:\epsilon$.

The unitary transformation to the doubly rotating frame used to derive Eq. \ref{eq:hamilt_stirap_2qubits}
is given by
\begin{equation}
    U_{x}(t)=e^{-i\br{\omega_s+\omega_p}t}\ketbra{1}{1}+e^{-i\omega_p t}\ketbra{2}
    {2}.
\end{equation}
The frequencies of the drives are taken as
\begin{subequations}
    \begin{align}
         & \omega_{p}= \varepsilon+ \frac{g}{2}- \Delta_{p}, \\
         & \omega_{s}= \varepsilon- \frac{g}{2}- \Delta_{s}.
    \end{align}
\end{subequations}
where $\Delta_{s}$ and $\Delta_{p}$ are the detunings. Denoting with $H(t)$ the
full Hamiltonian Eq.~\ref{eq:Hamiltonian_driven_nonoise} and transforming to the
rotating frame,
$\tilde{H}(t)=U_{x}(t)^{\dagger}H(t) U_{x}(t)-iU_{x}(t)^{\dagger}\partial_{t}U_{x}
(t)$, we obtain {\begin{multline}\label{H_rot1}\tilde{H}(t)=\br{\Delta_p+\Delta_s}\ketbra{1}{1}+ \Delta_{p}\ketbra{2}{2}+\br{\varepsilon+\frac{g}{2}}\ketbra{3}{3}+\\ +\left\{\alpha\left[ \frac{\Omega_{p}}{\sqrt{2}\alpha}\br{e^{-2i\omega_p t}+1}+\frac{\Omega_{s}}{\sqrt{2}\beta}\br{e ^{i\omega_{-}t}+e^{-i\omega_{+}t}}\right]\ketbra{0}{2}+\right.\\\left.+\beta\left[\frac{\Omega_{s}}{\sqrt{2}\beta}\br{e^{-2i\omega_s t}+1}+\frac{\Omega_{p}}{\sqrt{2}\alpha}\br{e ^{-i\omega_{-}t}+e^{-i\omega_{+}t}}\right]\ketbra{2}{1}\right.\\\left.+ h. c.\right\},\end{multline}}
where $\omega_{-}=\omega_{s}-\omega_{p}$ and $\omega_{+}=\omega_{s}+\omega_{p}$.
Since $2\omega_{s}, 2\omega_{p}, \omega_{+}\gg \omega_{s},\omega_{p}$ we can
neglect fastly oscillating terms\cite{ShoreFT1991theory}, obtaining
{\begin{multline}\label{H_rot2}\tilde{H}(t) = \Delta_{p}\ketbra{2}{2}+\br{\Delta_p+\Delta_s}\ketbra{1}{1}+\br{\varepsilon+\frac{g}{2}}\ketbra{3}{3}+\\ +\left\{\alpha\left[ \frac{\Omega_{p}}{\sqrt{2}\alpha}+\frac{\Omega_{s}}{\sqrt{2}\beta}e^{i\br{\Delta_s-\Delta_p-g}t}\right ]\ketbra{0}{2}+\right.\\\left.+\beta\left[\frac{\Omega_{s}}{\sqrt{2}\beta}+\frac{\Omega_{p}}{\sqrt{2}\alpha}e^{-i\br{\Delta_s-\Delta_p-g}t}\right]\ketbra{2}{1}+ h. c.\right\},\end{multline}}
If both the drives are resonant $\Delta_{p}=\Delta_{s}=0$ we obtain the
Hamiltonian eq.~(\ref{eq:hamilt_stirap_2qubits}) presenting a time-dependent term
oscillating at a frequency $g$.

\section{\label{sec:STIRAP}STIRAP}
STIRAP\cite{BergmannRMP1998coherent,VitanovRMP2017stimulated} is a protocol
yielding efficient and robust population transfer in a three-level system. We design
our system such as STIRAP is operated in $\mbox{span}\{\ket{0}, \ket{2}, \ket{1}\}$
(see Fig.~\ref{fig:system_energyscheme}). The system is driven by two time-dependent
classical fields: the \textit{pump pulse} $\Op(t)$ drives the transition $\ket{0}
-\ket{2}$, while the \textit{Stokes pulse} drives the transition $\ket{2}-\ket{1}$.
The Hamiltonian in an appropriate rotating frame reads~\cite{GiannelliPLA2022tutorial}
\begin{equation}
    \label{eq:stirap}
    \begin{aligned}
        H_{\mathrm{S}}(t)={} & \Delta\ketbra{1}{1}+\Delta_{p}\ketbra{2}{2}+{}                          \\
                             & {}+\frac{1}{2}\br{\Op(t)\ketbra{0}{2}+\Os(t)\ketbra{2}{1}+\text{h.c.}},
    \end{aligned}
\end{equation}
where $\Delta_{p}= \varepsilon + \frac{g}{2}- \omega_{\mathrm{p}}$ and
$\Delta = 2\varepsilon - \omega_{\mathrm{p}}-\omega_{\mathrm{s}}$ are the single
and two-photon detunings, respectively, and $\omega_{\mathrm{p}}$ and
$\omega_{\mathrm{s}}$ are the frequencies of the pump and Stokes fields,
respectively. For efficient population transfer, it is crucial to operate at
small two-photon detuning, $\Delta\ll\Omega_{\mathrm{p,s}}$. In particular, under
two-photon resonance ($\Delta=0$), $H_{\mathrm{S}}$ admits the so-called ``dark
state" as an instantaneous eigenstate
\begin{equation}
    \label{eq:darkstate}\ket{\phi_\mathrm{D}}=\cos{\theta(t)}\ket{0}-\sin{\theta(t)}
    \ket{1},
\end{equation}
with $\tan\theta(t) = \Op(t)/\Os(t)$. If the pulses are shined in a \textit{counterintuitive
sequence}, \ie, $\Os$ is applied before $\Op$, while ensuring that they overlap
for part of the protocol, then $\theta(t)$ smoothly varies from $0$ to $\pi/2$.
When the system is initialized in the state
$\ket{\phi_\mathrm{D}(\ti)}= \ket{0}$, if the evolution remains adiabatic~\cite{BornZFP1928beweis,Messiah1961quantum}
the system will follow the dark state $\ket{\phi_\mathrm{D}}$ reaching the desired
target state at the final time $\ket{\phi_\mathrm{D}(t_f)}=\ket{1}$.

It is well known that STIRAP is robust against variations in the pulse shapes~\cite{VitanovRMP2017stimulated}.
Among the many possible choices~\cite{GiannelliPRA2014threelevel}, in this work,
we use Gaussian pulses, given by
\begin{equation}
    \label{eq:Gaussian_pulses}\Op(t) = \Opmax e^{-\left(\frac{t-\tau}{T}\right)^2}
    , \quad \Os(t) = \Osmax e^{-\left(\frac{t+\tau}{T}\right)^2},
\end{equation}
and evolve the system over the time interval $[-5T,5T]$, with $\tau=0.7T$. Adiabaticity
is ensured by the global condition~\cite{BergmannRMP1998coherent}
\begin{equation}
    \label{eq:globaladiabaticitycondition}\Opsmax \tau \geq 10.
\end{equation}
The efficiency of the protocol is defined as the population of the target state $\ket
{1}$ at the end of the evolution. The formal definition is given directly in Sec.~\ref{sec:efficiency}
for the system of interest.

\section{Derivation of the Master Equation for Markovian noise}
\label{sec:LME} In this section we outline the derivation of the Master Equation
for Markovian (anti)correlated noise. The system Hamiltonian is defined by eq.~\ref{eq:Hamiltonian_driven_nonoise},
while the effect of the noise is described by eq. \ref{eq:Hamiltonian_noise}, with
$\delta_{1}=\frac{1}{\sqrt{\eta}}\chi$ and $\delta_{2}=\sqrt{\eta}\chi$,
obtaining: {\begin{equation}H_{n}=\chi(t)O,\end{equation}} with
\begin{equation}
    \begin{aligned}
        O = & {\frac{1}{2}\Bigg[\br{\frac{1+\eta}{\sqrt{\abs{\eta}}}\cos{\theta_e}}\br{\ketbra{1}{1}-\ketbra{0}{0}}+} \\
            & {-\br{\frac{g}{2\varepsilon}\frac{1+\eta}{\sqrt{\abs{\eta}}}}\br{\ketbra{0}{1}+\ketbra{1}{0}}}          \\
            & +{\br{\frac{1-\eta}{\sqrt{\abs{\eta}}}}\br{\ketbra{2}{3}+\ketbra{3}{2}}\Bigg]}.
    \end{aligned}
\end{equation}
We consider noise with zero mean, $\avg{\chi(t)}=0$, and Markovian $\avg{\chi(t)\chi(t^\prime)}
=\gamma \delta(t-t^{\prime})$. The density matrix $\tilde \rho(t)$ in the
interaction picture solves the von Neumann equation
$\dot{\tilde{\rho}}(t)=-i\left[\tilde{H}_{n},\tilde{\rho}(t)\right]$ which can
be formally integrated and averaged over the stochastic process yielding
\begin{equation}
    \label{eq:lme_avg}\avg{\dot{\tilde{\rho}}(t)}=-\avg{\int_0^t\left[\chi(t)\tilde{O}(t),\left[\chi(t^\prime)\tilde{O}(t^\prime),\tilde{\rho}(t^\prime)\right]\right]dt^\prime}
    ,
\end{equation}
For Markovian noise, we can write $\avg{\chi(t)\chi(t^\prime)\tilde{\rho}(t^\prime)}
=\avg{\chi(t)\chi(t^\prime)}\tilde{\rho}(t^{\prime})$, since $t^{\prime}<t$, and
$\tilde{\rho}(t^{\prime})$ cannot depend on $t$, except for $t^{\prime}=t$, obtaining
\begin{equation}
    \label{eq:lme_equation_instant}\avg{\dot{{\tilde{\rho}}}(t)}={\gamma}\br{{\tilde{O}^\dagger(t)}{\tilde{\rho}}(t)\tilde{O}(t)-\frac{1}{2}\left[\tilde{O}^\dagger(t)\tilde{O}(t),{\tilde{\rho}}(t)\right]_+}
\end{equation}
which in the Schrödinger picture reads
\begin{equation}
    \label{eq:lme_equation_appendix}\avg{\dot{{\rho}}(t)}=-i\left[H(t),\rho( t)\right
    ] +\gamma\br{{O^\dagger}{\rho}(t){O}-\frac{1}{2}\left[{O^\dagger}{O},{\rho}(t)\right]_+}
\end{equation}

{For zero mean, uncorrelated Markovian local noises, the noise Hamiltonian is \begin{equation}H_{n}(t) = \sum_{k = 1, 2}\delta_{k}(t)O_{k},\end{equation} the two collapse operators being \begin{align}O_{1} =&\frac{1}{2}\big[\cos{\theta_e}\big(\ketbra{1}{1}- \ketbra{0}{0}\big) + \big(\ketbra{2}{3}+\ketbra{3}{2}\big) \nonumber\\ +&\frac{g}{2\varepsilon}\ketbra{0}{1}+ h.c\big],\end{align} \begin{align}O_{2} =&\frac{1}{2}\big[\cos{\theta_e}\big(\ketbra{1}{1}- \ketbra{0}{0}\big) - \big( \ketbra{2}{3}+\ketbra{3}{2}\big) \nonumber\\&+\frac{g}{2\varepsilon}\ketbra{0}{1}+ h.c].\end{align} Repeating the same steps as before, and using the relation \begin{equation}\avg{\delta_k(t)\delta_l(t)}= \gamma_{k} \,\delta_{kl}\,\delta(t-t^{\prime}); \quad k, l = 1,2,\end{equation} cross terms vanish and in the Schrödinger picture we obtain \begin{equation}\begin{aligned}\label{eq:lme_equation_ucm}\avg{\dot{{\rho}}(t)}=&-i\left[H(t),\rho(t)\right] + \\&+ \sum_{k}{\gamma_k}\br{{O^\dagger_k}{\rho}(t){O_k}-\frac{1}{2} \big[{O^\dagger_k O_k},{\rho}(t)\big]_+}.\end{aligned}\end{equation} }
\section{\label{sec:NN}Neural Networks}
Neural networks (NNs)\cite{Goodfellow2016deep, bishop2006pattern} are at
the heart of modern artificial intelligence and have become essential in a wide
range of applications, including computer vision, speech recognition, natural language
processing, and autonomous systems. These models excel at capturing complex, non-linear
patterns in data, enabling machines to perform classification, regression, and decision-making
tasks with impressive accuracy. Inspired by the structure of the human brain, neural
networks are composed of layers of interconnected artificial neurons that learn representations
from data through iterative optimization techniques. Their flexibility, adaptability,
and scalability make them indispensable in both theoretical research and real-world
engineering solutions across diverse domains.

A NN is a function \cite{MarquardtSPLN2021machinea}
\begin{equation}
    \textrm{NN}_{\vect{\Theta}}(\vect{x}): \vect{x}\to\vect{y},\quad \vect{x}\in\mathcal{R}
    ^{n},\vect{y}\in\mathcal{R}^{m},
\end{equation}
which depends on \textit{many} parameters $\vect{\Theta}$. We employ a feedforward
neural network, specifically a multi-layer perceptron (MLP) with three hidden
layers in addition to the input and output layers. The detailed architecture is presented
in Table \ref{tab:NNstructure}.

A fundamental aspect of supervised learning is the requirement of a labeled dataset
for neural network training \cite{MarquardtSPLN2021machine}\cite{AIRFA2018handson}.
This includes pairs $(\vect{x}_{i},\vect{\hat y}_{i})_{i= 1,2,\dots,N}$, where
each input vector $\vect{x_i}$ has an associated output vector $\vect{\hat y_i}$.
The output from any given layer $(l)$ is represented as a vector $\vect{y}^{(l)}$,
where each component of this vector corresponds to the output generated by the
neurons in that layer
\begin{equation}
    \label{eq:yl}\vect{y}^{(l)}= f^{(l)}(\vect{z}^{(l)}),
\end{equation}
where, $f^{(l)}$ denotes a nonlinear function known as the \textit{activation
function} \cite{haykin2009neural, Geron2023handson}, and
\begin{equation}
    \label{eq:zl}\vect{z}^{(l)}= \vect{w}^{(l)}\vect{y}^{(l-1)}+ \vect{b}^{(l)}.
\end{equation}
Here, $\vect{w}^{(l)}\in \mathbb{R}^{D^{(l)} \times D^{(l-1)}}$ is the weight matrix,
$\vect{b}^{(l)}\in \mathbb{R}^{D^{(l)}}$ is the bias vector, $D^{(l)}$ represents
the dimension or number of neurons in $l$-th layer. Training the model begins by
randomly initializing the weights and biases, and then proceeds by updating them
using a variation of \textit{stochastic gradient descent}, in particular we used
the \textit{adam} optimizer~\cite{kingma2014adam, adam}. This process
stops when the cost function $C(\{\vect{y}_{i}, \vect{\hat{y}}_{i}\}_{i})$ is
minimized. We have opted for \textit{Sparse Categorical Crossentropy}~\cite{Geron2023handson}
as our cost function as mentioned in the main text. Although both \textit{Categorical
Crossentropy}\cite{Goodfellow2016deep} and \textit{Sparse Categorical
Crossentropy} share the same functional structure, the latter offers benefits due
to its use of integer labeling. This approach requires less memory and
computations, unlike \textit{Categorical Crossentropy}, which necessitates
one-hot encoding~\cite{bishop2006pattern, Goodfellow2016deep} for the
labels.

As shown in Table~\ref{tab:NNstructure}, the implementation includes \textit{leaky
rectified linear unit} (LeakyReLU)\cite{glorot2011deep} for the hidden
layers as the activation function to prevent the `dying ReLU' issue~\cite{glorot2011deep}
and achieve superior outcomes. The function is defined as:
\begin{equation}
    \label{eq:LeakyRELU}f_{\text{LReLU}}(z) =
    \begin{cases}
        z        & \text{if} \quad z \geq 0, \\
        \alpha z & \text{if} \quad z < 0,
    \end{cases}
\end{equation}
where $\alpha$ is set to $0.01$. In the output layer, the \textit{Softmax} activation~\cite{mackay2003information, bishop2006pattern}
function was employed to ensure that the resulting output from each neuron span
from $0$ to $1$ and adds up to $1$. These values indicate the probabilities that
the noise affecting the qubit falls into one of the $6$ categories
\begin{equation}
    \label{eq:softmax}\vect{y}= f_{\textrm{softmax}}(\vect{z}^{(L)}) = \frac{e^{\vect{z}^{(L)}}}{\sum_{k=1}^{D^{(L)}}e^{z_k^{(L)}}}
    ,
\end{equation}
where the exponentiation and division are performed element-wise.

\section*{Authors Declarations}
    The authors have no conflicts to disclose.
    
 \begin{acknowledgments}
    GF and LG acknowledge support from the PNRR MUR project PE0000023-NQSTI ``National
    Quantum Science and Technology Institute"; SM acknowledge support from the ICSC
    - Centro Nazionale di Ricerca in High-Performance Computing, Big Data and
    Quantum Computing; GF acknowledges support from PRIN 2022 ``SuperNISQ"; GF and
    EP acknowledge support from the University of Catania, Piano Incentivi
    Ricerca di Ateneo 2024-26, project QTCM; MP acknowledges funding from the
    Royal Society Wolfson Fellowship (RSWF/R3/183013), the Department for the Economy
    of Northern Ireland under the US-Ireland R\&D Partnership Programme, and the
    EU Horizon Europe EIC Pathfinder project QuCoM (GA no. 10032223). EP
    acknowledges the COST Action SUPERQUMAP (CA 21144).
\end{acknowledgments}

\FloatBarrier
\bibliography{TwoQubitsCorrelationsSTIRAPML}

\end{document}